\documentclass[pamm,a4paper,fleqn,oneside]{w-art}
\usepackage{times}
\usepackage{w-thm}

\theoremstyle{plain}

\theoremstyle{definition}

\usepackage[]{graphicx}
\chardef\bslash=`\\ % p. 424, TeXbook

\newcommand{\ud}{\mathrm{d}} 
\begin{document}
%%    \renewcommand{\copyrightyear}{2004}
%%    \DOIsuffix{pamm.20041zzzz}
%%    \Volume{4} \Year{2004} \pagespan{xxx}{yyy}

%%%%%%%%%%%%%%%%%%%%%%%%%%%%%%%%%%%%%%%%%%%%%%%%%%%%%%%%%%%%%%%%%%%%%%%%%%
\begin{minipage}[t]{180mm}
\thispagestyle{empty}
\vspace{20mm}

\begin{center}
{\Large\bf Localization and Pattern Formation}

\vspace{7mm}

{\Large\bf in BBGKY Hierarchy} 

\vspace{20mm}

{\large\bf Antonina N. Fedorova, Michael G. Zeitlin}

\vspace{20mm}

Mathematical Methods in Mechanics Group \\

Institute of Problems of Mechanical Engineering (IPME RAS)\\ 

Russian Academy of Sciences \\

Russia, 199178, St. Petersburg, V.O., Bolshoj pr., 61\\

zeitlin@math.ipme.ru, anton@math.ipme.ru\\
         
http://www.ipme.ru/zeitlin.html\\

http://www.ipme.nw.ru/zeitlin.html

\vspace{20mm}
{\bf Abstract}

\vspace{10mm}

\begin{tabular}{p{100mm}}
A fast and efficient numerical-analytical approach
is proposed for modeling complex behaviour
in the BBGKY hierarchy  of kinetic equations.
Numerical modeling shows the creation of
various internal
structures from localized modes, which are related to the localized or chaotic type
of behaviour and
the corresponding patterns (waveletons) formation.

\vspace{10mm}

Presented: GAMM Meeting, 2004, Dresden, Germany.
\vspace{5mm}

Published: Proc. Appl. Math. Mech. (PAMM), {\bf 4}, 564, Wiley-VCH, 2004.

\end{tabular}

\end{center}
\end{minipage}
\newpage

%%%%%%%%%%%%%%%%%%%%%%%%%%%%%%%%%%%%%%%%%%%%%%%%%%%%%%%%%%%%%%%%%%%%%%%%%%%%%%%%%%%%

\title{Localization and Pattern Formation in BBGKY Hierarchy}
\author{Antonina N. Fedorova}
\author{Michael G. Zeitlin\footnote{Corresponding
     author: e-mail: {\sf zeitlin@math.ipme.ru}, http://www.ipme.ru/zeitlin.html,
  http://www.ipme.nw.ru/zeitlin.html}}
\address[]{IPME RAS, St.~Petersburg, V.O. Bolshoj pr., 61, 199178, Russia}

\begin{abstract} 
A fast and efficient numerical-analytical approach
is proposed for modeling complex behaviour
in the BBGKY hierarchy  of kinetic equations.
Numerical modeling shows the creation of
various internal
structures from localized modes, which are related to the localized or chaotic type
of behaviour and
the corresponding patterns (waveletons) formation. 

\vspace{1pc}
\end{abstract}

\maketitle

We consider the application of a new nu\-me\-ri\-cal/analytical 
technique based on local nonlinear harmonic analysis 
approach (multiscale decomposition) for 
the description of complex (non-equilibrium) behaviour  
of the classical statistical ensembles, considered in the framework 
of the general BBGKY hierarchy (kinetics equations). 
We restrict ourselves to the rational/polynomial type of
nonlinearities (with respect to the set of all dynamical variables)
that allows  to use our results from 
[1],[2], which are based on the so called multiresolution
framework [3] and the variational formulation of initial nonlinear (pseudodifferential) problems.
Our main goals are an attempt of classification and construction 
of a possible zoo of nontrivial (meta) stable states:
high-localized (nonlinear) eigenmodes, complex (chaotic-like or entangled) 
patterns, localized (stable) patterns (waveletons). 
The last case 
is a good image for plasma modeling in  fusion state (energy confinement).
It should be noted that the class of smoothness (related at least to the appearance
of chaotic/fractal-like type of behaviour) of the proper functional space 
under consideration 
plays a key role in the following [2].
Let $M$ be the phase space of an ensemble of $N$ particles ($ {\rm dim}M=6N$)
with coordinates
$x_i=(q_i,p_i), \quad i=1,...,N, \quad
q_i=(q^1_i,q^2_i,q^3_i)\in R^3,\quad
p_i=(p^1_i,p^2_i,p^3_i)\in R^3,\quad
q=(q_1,\dots,q_N)\in R^{3N}$.
Individual and collective measures are: 
$
\mu_i=\ud x_i=\ud q_i\ud p_i,\quad \mu=\prod^N_{i=1}\mu_i
$.
Our constructions can be applied to the following general Hamiltonians:
$
H_N=
\sum^N_{i=1}\Big(\frac{p^2_i}{2m}+U_i(q)\Big)+
\sum_{1\leq i\leq j\leq N}U_{ij}(q_i,q_j)  
$,
where the potentials 
$U_i(q)=U_i(q_1,\dots,q_N)$ and $U_{ij}(q_i,q_j)$
are restricted to rational functions of the coordinates.
Let $L_s$ and $L_{ij}$ be the Liouvillean operators (vector fields) 
and
$
F_N(x_1,\dots,x_N;t)
$
be the hierarchy of $N$-particle distribution function,
satisfying the standard BBGKY--hierarchy ($V$  is the volume):
\begin{eqnarray}
\frac{\partial F_s}{\partial t}+L_sF_s=
\frac{1}{V^s}\int\ud\mu_{s+1}
\sum^s_{i=1}L_{i,s+1}F_{s+1}
\end{eqnarray}
Our key point in the following consideration is the proper nonperturbative 
generalization of the perturbative multiscale approach of Bogolyubov.
The infinite hierarchy of distribution functions satisfying system (1)
in the thermodynamical limit is:
$
F=\{F_0,F_1(x_1;t),\dots,
F_N(x_1,\dots,x_N;t),\dots\}
$,
where
$F_p(x_1,\dots, x_p;t)\in H^p$,
$H^0=R,\quad H^p=L^2(R^{6p})$ (or any different proper functional spa\-ce), $F\in$
$H^\infty=H^0\oplus H^1\oplus\dots\oplus H^p\oplus\dots$
with the natural Fock space like norm 
(guaranteeing the positivity of the full measure):
\begin{eqnarray}
(F,F)=F^2_0+\sum_{i}\int F^2_i(x_1,\dots,x_i;t)\prod^i_{\ell=1}\mu_\ell.
\end{eqnarray}
First of all we consider $F=F(t)$ as a function of time only,
$F\in L^2(R)$, via
multiresolution decomposition which naturally and efficiently introduces 
the infinite sequence of the underlying hidden scales.
Because the affine
group  of translations and dilations 
generates multiresolution approach, this
method resembles the action of a microscope. 
We consider a multiresolution decomposition of $L^2(R)$ [3]
(of course, we may consider any different and proper for some particular case functional space)
which is a sequence of increasing closed subspaces $V_j\in L^2(R)$ 
(subspaces for 
modes with fixed dilation value):
$
...V_{-2}\subset V_{-1}\subset V_0\subset V_{1}\subset V_{2}\subset ...
$.
The closed subspace
$V_j (j\in {\bf Z})$ corresponds to  the level $j$ of resolution, 
or to the scale j
and satisfies
the following properties:
let $W_j$ be the orthonormal complement of $V_j$ with respect to $V_{j+1}$: 
$
V_{j+1}=V_j\bigoplus W_j.
$
Then we have the following decomposition:
$
\{F(t)\}=\bigoplus_{-\infty<j<\infty} W_j =
\overline{V_c\displaystyle\bigoplus^\infty_{j=0} W_j}
$
in case when $V_c$ is the coarsest scale of resolution.
The subgroup of translations generates a basis for the fixed scale number:
$
{\rm span}_{k\in Z}\{2^{j/2}\Psi(2^jt-k)\}=W_j.
$
The whole basis is generated by action of the full affine group:
$
{\rm span}_{k\in Z, j\in Z}\{2^{j/2}\Psi(2^jt-k)\}=
{\rm span}_{k,j\in Z}\{\Psi_{j,k}\}
=\{F(t)\}
$. In multidimensional case we may consider polynomial tensor bases, e.g. in $n=2$ case
we may use the rectangle lattice of scales and one-dimensional wavelet
decomposition:
$
f(x_1,x_2)=\sum_{i,\ell;j,k}\langle f,\Psi_{i,\ell}\otimes\Psi_{j,k}\rangle
\Psi_{j,\ell}\otimes\Psi_{j,k}(x_1,x_2)
$,
where the basis functions $\Psi_{i,\ell}\otimes\Psi_{j,k}$ depend on
two scales $2^{-i}$ and $2^{-j}$.
We obtain our multiscale\-/mul\-ti\-re\-so\-lu\-ti\-on 
representations (formulae (3) below) 
via the variational wavelet approach for 
the following formal representation of the BBGKY system (1) 
(or its finite-dimensional nonlinear approximation for the 
$n$-particle distribution functions) 
with the
corresponding obvious constraints on 
the distribution functions.
Let $L$ be an arbitrary (non)li\-ne\-ar dif\-fe\-ren\-ti\-al\-/\-in\-teg\-ral operator 
 with matrix dimension $d$
(finite or infinite), 
which acts on some set of functions
from $L^2(\Omega^{\otimes^n})$:  
$\quad\Psi\equiv\Psi(t,x_1,x_2,\dots)=\Big(\Psi^1(t,x_1,x_2,\dots), \dots$,
$\Psi^d(t,x_1,x_2,\dots)\Big)$,
 $\quad x_i\in\Omega\subset{\bf R}^6$, $n$ is the number of particles:
$
L\Psi\equiv L(Q,t,x_i)\Psi(t,x_i)=0$,
$Q\equiv Q_{d_0,d_1,d_2,\dots}(t,x_1,x_2,\dots$,
$\partial /\partial t,\partial /\partial x_1$,
$\partial /\partial x_2,\dots,\int \mu_k)=
\sum_{i_0,i_1,i_2,\dots=1}^{d_0,d_1,d_2,\dots}$
$q_{i_0i_1i_2\dots}(t,x_1,x_2,\dots)$
$\Big(\frac{\partial}{\partial t}\Big)^{i_0}\Big(\frac{\partial}{\partial x_1}\Big)^{i_1}$
$\Big(\frac{\partial}{\partial x_2}\Big)^{i_2}\dots\int\mu_k
$.
Let us consider  the $N$ mode approximation for the solution as 
the following ansatz:
$
\Psi^N(t,x_1,x_2,\dots)=
\sum^N_{i_0,i_1,i_2,\dots=1}a_{i_0i_1i_2\dots}
 A_{i_0}\otimes 
B_{i_1}\otimes C_{i_2}\dots(t,x_1,x_2,\dots)
$.
We shall determine the expansion coefficients from the following conditions:
$
\ell^N_{k_0,k_1,k_2,\dots}\equiv 
\int(L\Psi^N)A_{k_0}(t)B_{k_1}(x_1)C_{k_2}(x_2)\ud t\ud x_1\ud x_2\dots=0
$.
Thus, we have exactly $dN^n$ algebraical equations for  $dN^n$ unknowns 
$a_{i_0,i_1,\dots}$.
This variational ap\-proach reduces the initial problem 
to the problem of solution 
of functional equations at the first stage and 
some algebraical problems at the second.
The solution is parametrized by the solutions of two sets of 
reduced algebraical
problems, one is linear or nonlinear
(depending on the structure of the operator $L$) and the rest are linear
problems related to the computation of the coefficients of the algebraic equations.
The solution of the equations (1) has the 
following mul\-ti\-sca\-le decomposition via 
high\--lo\-ca\-li\-zed eigenmodes
\begin{eqnarray}
&&F(t,x_1,x_2,\dots)=
\sum_{(i,j)\in Z^2}a_{ij}U^i\otimes V^j(t,x_1,\dots),\\
&&V^j(t)=
V_N^{j,slow}(t)+\sum_{l\geq N}V^j_l(\omega_lt), \ \omega_l\sim 2^l,\ 
U^i(x_s)=
U_M^{i,slow}(x_s)+\sum_{m\geq M}U^i_m(k^{s}_mx_s), \ k^{s}_m\sim 2^m,\nonumber
\end{eqnarray}
which corresponds to the full multiresolution expansion in all underlying time/space 
scales.
The formulae (3) give the expansion into a slow 
and fast oscillating parts.  So, we may move
from the coarse scales of resolution to the 
finest ones for obtaining more detailed information about the dynamical process.
In this way one obtains contributions to the full solution
from each scale of resolution or each time/space scale or from each nonlinear eigenmode.
It should be noted that such representations 
give the best possible localization
properties in the corresponding (phase)space/time coordinates. 
Formulae (3) do not use perturbation
techniques or linearization procedures.
Numerical calculations are based on compactly supported
wavelets and related wavelet families [3] and on evaluation of the 
accuracy on 
the level $N$ of the corresponding cut-off of the full system (1) 
regarding norm (2):
$
\|F^{N+1}-F^{N}\|\leq\varepsilon.
$
Fig.~1 demonstrates the appearance of localized (meta) stable pattern (waveleton),
which can be considered, e.g., as a model for fusion state in plasma.

\begin{vchfigure}[htb]
\includegraphics[width=.5\textwidth]{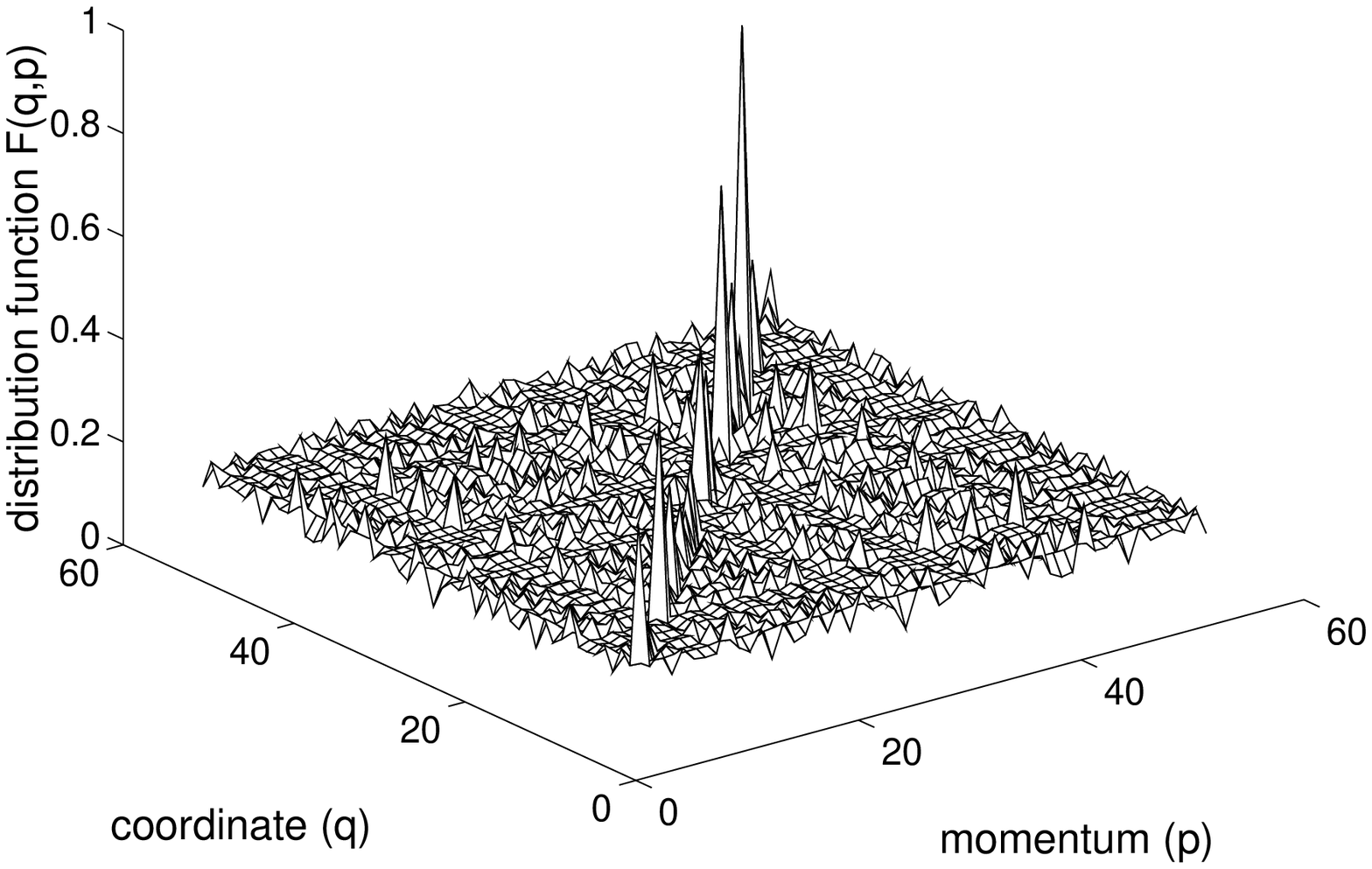}
\vchcaption{Localized waveleton pattern.}
\end{vchfigure}

\vspace{\baselineskip}

\end{document}